\documentclass[preprint,aps,showpacs,floatfix]{revtex4}
\usepackage{graphicx}
\usepackage{graphics}
\usepackage{color}      
\usepackage{subfigure}

\newcommand{\etal}{{\it et al.}~}
\newcommand{\etalcc}{{\it et al.}}

\begin{document}


\definecolor{AsGreen}{rgb}{0.3,0.8,0.3} 
\definecolor{AsRed}{rgb}{0.6,0.0,0.0}
\definecolor{AsLightBlue}{rgb}{0,0.6,0.6}
\definecolor{AsPurple}{rgb}{0.6,0,0.6}

\hyphenpenalty=5000
\tolerance=1000

\title{Non-classical velocity statistics in a turbulent atomic 
Bose--Einstein condensate}
\author{Angela White}
\email{ang.c.white@gmail.com}
\author{Nick P. Proukakis}
\author{Carlo F. Barenghi}
\affiliation{School of Mathematics and Statistics, 
Newcastle University, Newcastle upon Tyne, NE1 7RU, England, UK.}
\date{\today}
\begin{abstract}
In a recent experiment
Paoletti {\it et al.} (Phys. Rev. Lett. {\bf 101}, 154501 (2008))
monitored the motion of tracer particles in turbulent superfluid helium
and inferred that the velocity components do not obey the 
Gaussian statistics observed in ordinary turbulence.
Motivated by their experiment, we create a small turbulent state 
in an atomic Bose-Einstein condensate, which enables us to
compute directly the velocity field, and we find similar 
non-classical power-law tails. 
Our result thus suggests that non-Gaussian turbulent velocity
statistics describe a fundamental property of quantum fluids.
We also track the decay of the vortex tangle in the presence of
the thermal cloud.
\end{abstract}
\maketitle

Quantum turbulence (a dynamic tangle of discrete, 
reconnecting vortices) has been studied in
superfluid $^4$He \cite{turbo-he4}, 
superfluid $^3$He-B \cite{turbo-he3}, 
and, more recently, in atomic Bose--Einstein 
condensates (BEC's) \cite{Horng09,Tsubota-review,Bagnato}.
The defining property of these quantum fluids is that the 
superfluid velocity field is proportional to the gradient of the
phase of a complex order parameter, so the circulation is
quantized in unit of the quantum $\kappa$ (the ratio of Planck
constant to the mass of the relevant boson). Therefore, whereas in 
classical ordinary fluids (such as air  or water) the rotational
motion is unconstrained, the velocity field around a quantum vortex
decreases strictly as $\kappa/(2\pi r)$ where $r$ is the distance 
away from the vortex axis.   

Recent work has revealed that there are remarkable similarities between 
classical turbulence and quantum
turbulence: the same Kolmogorov energy spectrum
in continuously excited turbulence  \cite{spectrum},
the same temporal decay of the vorticity in decaying turbulence
\cite{decay}, the same pressure drops in pipe and channel 
flows \cite{pipe}, and the same drag crisis behind a sphere 
moving at high velocity \cite{sphere}. 
These results have attracted attention to the problem of the
relation between a classical fluid (which obeys the Euler
or Navier--Stokes equations) 
and a quantum fluid \cite{classical-quantum},
stimulating apparently simple questions such as whether
classical behaviour is merely the consequence of many quanta.
A recent experiment by Paoletti \etal \cite{Paoletti} 
in superfluid $^4$He has shed light onto this problem.
Using a new visualization technique based on solid hydrogen tracers, 
Paoletti \etal found that the components of the turbulent
velocity field do not obey the usual Gaussian distribution which is
observed in ordinary turbulence \cite{classical-turbo},
but rather follow power--law like behaviour.

The main aim of this Letter is to answer the question of
whether non-Gaussian velocity statistics is a fundamental
property of turbulence in a quantum fluid. To achieve this aim,
we move from $^4$He to an atomic BEC. The agreement that we find
between these two distinct systems
means that power--law behaviour is indeed typical of 
quantum fluids, in stark contrast with classical turbulence. 

To create a vortex tangle in a harmonically confined
atomic BEC we choose the technique of phase 
imprinting \cite{phase-imprinting}.
Although phase imprinting is not the only method which can be used
to generate turbulence \cite{createturb,Bagnato}, it is specific to 
BEC's, and can in principle be exploited to non-destructively 
generate tangles engineered to be isotropic or polarized; this
gives the technique its own numerical advantages over others.
Having generated a vortex tangle, we also study its decay 
at zero temperature by evolving the three--dimensional (3D) 
Gross--Pitaevskii equation for realistic experimental parameters, 
and also discuss the effect of the thermal cloud
on this decay.

An ultra--cold dilute atomic gas confined in a spherical trap is
described by the Gross--Pitaevskii equation (GPE) for the complex
wavefunction $\Psi$. It is convenient to consider the GPE
in dimensionless form as
\begin{equation}\label{dimgpe}
i \frac{\partial}{\partial t} {\psi}
  =\left(-\frac{1}{2} \bigtriangledown^{2}+\frac{1}{2} \textbf{r}^{2}
+ C\left| {\psi}  \right|^{2} \right){\psi}  \, ,
\end{equation}
\noindent where $\textbf{r}$ is the position vector; 
 ${\psi}= a_{o}^{3/2} N^{-1/2} \Psi $ is dimensionless and satisfies the
constraint $\int \text{d\textbf{r}}\left| {\psi} \right|^{2} = 1$. 
In writing Eq.\ (\ref{dimgpe}) we use the harmonic oscillator
length $a_{o}=\sqrt{\hbar / (m \omega)}$ 
as the unit of distance and the inverse trapping frequency
$1/\omega$ as the unit of time, where $m$ is the mass of one atom, 
$C =  4\pi N a / a_{o}$ is the dimensionless measure of the interaction 
between bosons, $N$ is the number of atoms and $a$ is the scattering length.

Finite temperature effects can be simulated by replacing
$i$ at the L.H.S. of Eq.\ (\ref{dimgpe}) with $(i-\gamma)$, where
$\gamma$ models the dissipation arising from the thermal cloud.
Although such a model can be justified from first principles 
\cite{gamma_origin}, one often resorts to the use of a 
(phenomenological) constant dissipation term 
\cite{gamma} (as done here), as it is much less computationally 
intensive while still 
giving a qualitatively accurate prediction of the relevant physics.

\begin{figure}[!h]
\subfigure[]
{
\centering{
\includegraphics[width=0.3\textwidth]{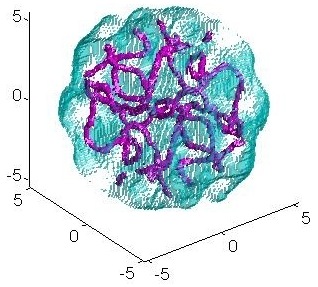}
}
}


\subfigure[]
{
\includegraphics[width=7cm]{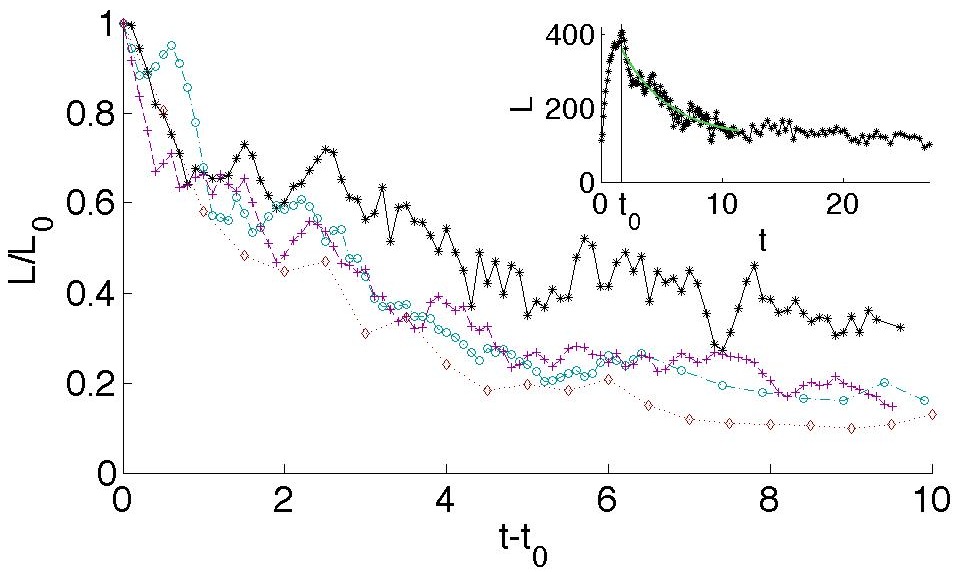}
}
\caption{
(Color online) 
(a): Generated turbulent state within the 
condensate edge (blue shading),
at $t_0=1.9$ and $\gamma = 0$; the maximum density (used
to determine the condensate edge) is $n=4.388 \times 10^{-3}$.
(b) Evolution of normalized total length $L(t-t_0)/L_{0}$ for 
$\gamma = 0$ (black asterixes, {\color{black}{$\ast$}})
$0.015$ (purple pluses, {\color{AsPurple}{$+$}}),
$0.03$ (blue circles, {\color{AsLightBlue}{$\circ$}})
and $0.06$ (red diamonds) corresponding to
the same initial phase imprinting on a $128^3$ grid;
the inset also shows the initial increase of $L$ 
for $t < t_0$ and the exponential fit for $\gamma=0$ (solid green line).  
}
\label{fig1}
\end{figure}

We employ realistic experimental parameters for a 
$^{23}$Na condensate ($a=2.75~\rm nm$) of 
$N=10^{5}$ atoms and trapping 
frequency $\omega = 2 \pi \times 150~\rm Hz$, 
giving $C = 2.019\times10^{3}$,  
$a_{o}=1.71~\mu \rm m$ and $1.06~\rm ms$ as time unit. 
In a homogeneous condensate the healing length $\xi$ is 
estimated by balancing the kinetic energy per particle and the 
interaction strength, which implies the (dimensionless)
$\xi =  (2 C n )^{-1/2}$
where $n=\vert \psi\vert^2$ is the  
condensate density. In a harmonic trap,  $n$ is position dependent, so
we estimate the average $\xi$ using the mean density; for example
at $t=t_0$ (Fig.~\ref{fig1}), 
$\langle n \rangle=1.35 \times 10^{-3}$ and $\langle \xi \rangle =0.43$.

Eq.\ (\ref{dimgpe}) is evolved pseudo-spectrally via XMDS \cite{XMDSdoc}
in 3D using a $4^{{\rm th}}$ order  Runge--Kutta method  
and periodic boundary conditions.
The spatial domain 
$|x|, |y|, |z| \leq 8$ is discretised on a $128\times128\times128$ grid 
with timestep $\Delta t=10^{-4}$. We have verified that 
the results are independent of spatial and temporal stepsize.

The initial condition is created by phase--imprinting a grid of 16 
vortices oriented parallel to the Cartesian axes in a staggered way (no
vortices intersect), and propagating this configuration for a short period 
in imaginary time
(which is equivalent to substituting $t$ with $-it$ in Eq.\ (\ref{dimgpe})) 
while continuously re-normalizing $\psi$, until the 
density adjusts to reveal the desired vortex structure. 
This is then propagated in real time.
In the initial evolution period,
the vortices reconnect and become excited, resulting in an increase
in the total vortex line length $L$.
This increase continues up to some time $t_0$, when 
the $L(t)$ achieves its peak value $L_{0}$,
and a maximally tangled turbulent state has been generated
(Fig.~\ref{fig1}(a)). At this point the dimensionless vortex
line density (vortex length divided by volume) is $0.79$.
The vortex line length evolution for $t>t_0$
is shown in Fig.~\ref{fig1}(b).
Since no energy is injected into the
system the turbulence is not sustained but decays over a time scale
of the order $t \approx 10$, as vortices further reconnect,
break up and decay into sound waves \cite{vortices-sound}, 
or leave the condensate. 
Our calculation lasts longer (up to $t=27$)
at which time one long vortex is left at the centre of the 
condensate and some shorter vortices, roughly aligned in the same 
direction, surround it closer to the condensate's edge.  
Large volume oscillations of the condensate at $1/3$ of the trap
frequency are observed.
Although the amplitude of these oscillations 
decreases as the turbulence decays, a method of exciting 
turbulence without inducing volume oscillations would 
be preferable for experiments. 

At each time $t$ we attribute a vortex core length 
to every point in the condensate at which the real and imaginary 
parts of $\psi$ crosses zero (therefore defining the vortex axis) \cite{Thesis}. 
The total vortex length $L(t)$ is the sum of all 
identified vortex points within the 
condensate edge, which, throughout this work, is 
defined as the outermost points at which $n$
drops below 25\% of the maximum density.
We find the decay of the vortex length is better fit by an exponential
of the form $L(t-t_0)/L_{0} \sim \exp(-c(t-t_0))$ (solid line in Fig.\ 1(b)), 
rather than $1/L(t-t_0)\propto t-t_0$ found in \cite{turbo-he4}.
Fitting over $10$ time units, yields 
consistent values for different grid sizes ($c=0.151\pm0.005$ 
and  $0.144\pm0.007$
for gridsizes of $128^{3}$ and $256^{3}$ respectively, with
corresponding line lengths $L_{0} = 406$ and $378$).
We have also checked that the decay of $L(t)$ is largely insensitive to the 
initial imprinted vortex configuration obtained 
by imprinting extra vortices.

To obtain a qualitative understanding of the effect of 
the thermal cloud, we have repeated the above calculation 
with dissipation parameters $\gamma=0.015,\,0.03$ and $0.06$. 
As anticipated, the role of temperature is to induce faster 
decay of the turbulent state, leading respectively to 
$c=0.252\pm0.007,\,0.274\pm0.006$ and $0.340\pm0.018$ 
(with corresponding $L_{0}=369,\,339,$ and $300$, 
due to the damping in the initial period $t < t_0$).

Our next step is to analyse the turbulent (dimensionless)
superfluid velocity field, which we compute directly from the definition
$\textbf{v}(\textbf{r})=\left({\psi}^{*}\bigtriangledown{\psi}-
{\psi}\bigtriangledown{\psi}^{*}\right) / (2 i \vert {\psi} \vert^{2})$.
%
%
\noindent (the derivatives of $\psi$ being obtained spectrally). The calculation
is tested against the expected azimuthal velocity profile of a single vortex
set along the $z$ axis, which is
$v_{\hat{\theta}} =\kappa / (2\pi r)$ where $\kappa = 2 \pi$ is 
the quantum of circulation.  
Using the Madelung representation 
${\psi} = \sqrt{n(\textbf{r})}\exp[i\varphi(\textbf{r})]$,
yields
$\textbf{v}(\textbf{r})=\nabla\varphi(\textbf{r})$, showing that
the velocity depends only on the condensate phase $\varphi$.

\begin{figure}[!h]
\includegraphics[angle=0,width=0.43\textwidth]{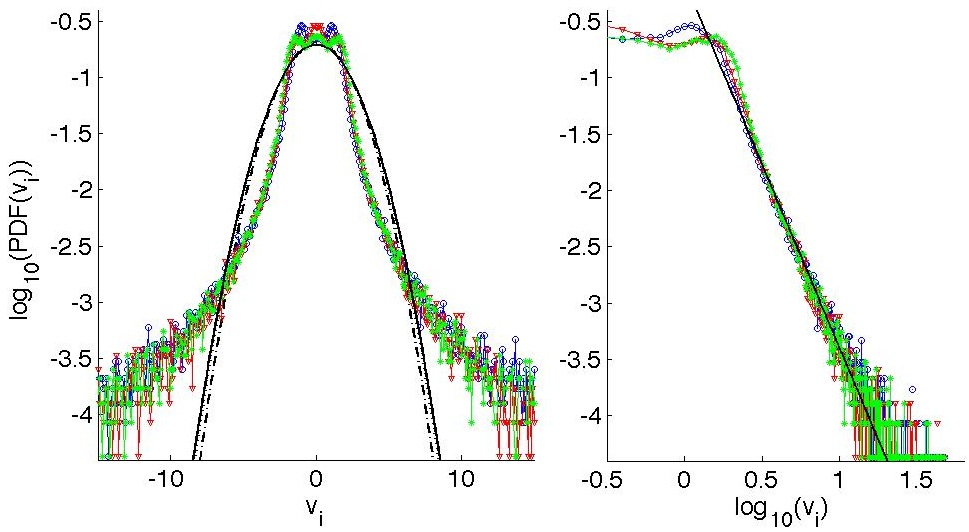}
\caption{
(Color online)
Left:
3D log-linear velocity PDFs corresponding to Fig.\ 1 ($\gamma=0$):
$\log_{10}[\text{PDF}(v_{i})]$ vs. velocity component $v_{i}$,
for $v_{x}$ (blue circles, {\color{blue}{$\circ$}}), 
$v_{y}$ (red triangles, {\color{red}{$\bigtriangledown$}})
and $v_{z}$ (green asterixes, {\color{green}{$\ast$}}) , yielding
following power law coefficients respectively
$b = -3.27\pm0.04,\,\,-3.54\pm0.06,\,\,-3.57\pm0.06$.
Corresponding $\log_{10}[\text{gPDF}(v_{i})]$ plots shown by
black dotted ($\cdots$, $v_{x}$), 
dash-dotted (-$.$, $v_{y}$) and uniform (--, $v_{z}$) lines,
which almost overlap.
The velocity components are only sampled within the condensate edge 
(defined at 25\% of the maximum density) thus excluding the 
outer region where the density drops to zero. 
Corresponding statistical values:
$\sigma^{2}_{i}=4.1,\,\,3.6,\,\,4.2$ and
 $\tilde{\mu}_{i}=\left(4.6,\,\,4.5,\,\,4.6\right)\times10^{-2}$. 
Right:
Log-log plot of the same PDFs; the solid line
is the power law fit to $v_x$.
} 
\label{fig2}
\end{figure}

We calculate the probability density function (PDF) of each  
Cartesian velocity component $v_i$ ($i=x,y,z$) for $\gamma=0$ and compare it
to the form that a Gaussian PDF (gPDF) of the velocity would take:
\begin{equation}\label{gpdf}
\text{gPDF}(v_{i})=\frac{1}{\sigma \sqrt{2\pi}}\exp\left(\frac{-(v_{i}-\tilde{\mu})^2}{2\sigma^{2}}\right)\,,
\end{equation}
\noindent
where $\sigma$, $\sigma^{2}$ and $\tilde{\mu}$ are the 
standard deviation, variance and mean of the velocity distribution (see Fig.~\ref{fig2}).
In evaluating the gPDF, we use the mean, standard deviation 
and variance of the actual velocity PDF. Power-law dependence of PDF($v_{i}$)$\propto v_{i}^{b}$ is found from $\log(\text{PDF})$ vs $\log(v_{i})$ plots of the positive velocity components, with $-3.57 < b < -3.27$ in all three directions.
%

It is apparent that our velocity statistics 
are non-Gaussian, consistently with the high velocity tails 
found experimentally by Paoletti \etalcc ~\cite{Paoletti} 
in turbulent superfluid $^4$He (their $b$ is $-3$, slightly
less than ours). 
It is well--known that the PDF's of velocity components in
ordinary turbulence are Gaussian
\cite{classical-turbo}; our result thus supports the idea
(put forward in Ref.~\cite{Paoletti})
that non--Gaussian velocity statistics are the 
unique signature of the quantized nature of the vortices. 

We find that deviations from Gaussian behaviour are less pronounced 
(but still noticeable) if we omit sampling the velocity very near 
the axis when the density is less than a prescribed cutoff. 
Unlike a classical vortex, there are not individually 
distinguishable atoms which spin about the vortex axis, and the
velocity field is entirely defined by the macroscopic 
phase $\varphi({\bf r})$ 
irrespective of the density $n({\bf r})$, 
so our procedure is justified.
The small wiggles in the PDF for values close to zero may 
be a measure
of the anisotropy of the vortex tangle in our 
condensate and would not feature in velocity PDFs  of a truly 
isotropic state of quantum turbulence in a larger condensate, 
or they could be due to the large volume 
oscillations of the 3D condensate.   

The observed non-Gaussianity of the velocity statistics 
holds during the decay of the vortex tangle, 
suggesting that it is not necessarily caused by 
vortex reconnections, whose frequency depends on the vortex line 
density \cite{reconnection-rate}. To verify that our result is general
and does not depend on vortex line density or reconnections,
we repeat the calculation in a two--dimensional (2D) BEC; 
such a condensate is created
when the axial trapping frequency, $\omega_{z}$, is much greater than 
the radial trapping frequency, $\omega_{r}$, freezing out motion 
along $z$.  The condensate in the radial plane is also described by 
Eq.\ (\ref{dimgpe}), with 
$C\rightarrow C_{2d}=2\sqrt{2\pi}aN/a_{z}$, 
and $\psi \rightarrow \psi_{2d}=a_{r}N^{-1/2} \Psi$ which 
relates the dimensionless
 wavefunction to the dimensional 2D condensate wavefunction, $\Psi_{2d}$,
and $a_{r}=\sqrt{\hbar/(m\omega_{r})}$ (here
$a_{z}=\sqrt{\hbar/(m\omega_{z})}$ are the harmonic oscillator lengths 
in the radial and $z$ directions).
To induce turbulence in a 2D BEC we imprint equal 
numbers of oppositely charged vortices randomly located and aligned 
along the axial direction \cite{Horng09}.
Here we simulate a 2D $^{23}$Na Bose gas with
$N=10^{7}$ atoms 
and condensate aspect ratio $\omega_{z}/\omega_{r}=20$ 
(with $\omega_{r}=7.5 \times 2 \pi~\rm Hz$), 
implying $C_{2d}=8.06\times10^{4}$.   
The spatial computational domain $ |x|, |y| \leq 25$ is 
discretized on a $1024 \times 1024$ grid with 
timestep $\Delta t=10^{-4}$. 
Choosing our initial monitoring time arbitrarily at $t_0=4.4$, 
we have 86 vortices in the dimensionless condensate area $752$,
corresponding to a dimensionless vortex line density (number of
vortices per unit area) of $0.11$.
We find a mean radial density $\langle n_{2d} \rangle =1.23\times10^{-4}$, 
and an estimated healing length 
$\xi_{2d}=(2C_{2d}n_{2d})^{-1/2}=0.225$. 
Proceeding as in 3D,
the velocity PDF's and gPDF's of the
2D condensate are shown in Fig.\ 3 for $\gamma=0$;
we find again
non-classical, non-Gaussian velocity statistics, with high velocity 
tails.  This PDF lacks the small wiggles 
close to zero velocity observed in the 3D
case, as the method of 
phase imprinting $42$ positive and $42$ negative vortices along the 
$z$ axis in randomly selected locations across a much larger 
condensate generates a more isotropic turbulent state with negligible oscillations of the total radial area. 
As in 3D, we find a power-law correlation with velocity, 
$\text{PDF}(v_{i})\propto v_{i}^{b}$ with 
$-3.18 \pm 0.08$ 
for positive $v_{x}$ and $v_{y}$ velocity components respectively. 
Unlike the 3D case, vortex reconnections have played a negligible
role in the state which is measured because during the measured
timescale the number of vortices
remains approximately the same as initially imprinted.

\begin{figure}[t]
\includegraphics[angle=0,width=0.48\textwidth]{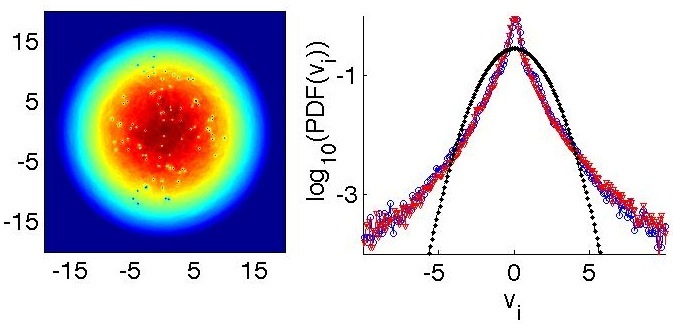}
\caption{
Density (left) and velocity PDFs (right) for a  2D BEC at 
$t_0=4.4$ and $\gamma=0$. The maximum density is
$n=2.057 \times 10^{-3}$.
Left: $86$ vortices (core radius $\approx 2.66$) can be identified.
Right: Plots of $\log_{10}[\text{PDF}(v_{i})]$ and the corresponding 
$\log_{10}[\text{gPDF}(v_{i})]$ vs $v_{i}$, where the 
$v_{x}$, $v_{y}$  velocity components (sampled within the condensate edge)
are given by 
blue circles ({\color{blue}{$\circ$}}) and 
red triangles ({\color{red}{$\bigtriangledown$}}); 
 corresponding  gPDF's displayed by black dotted ($\cdots$) and 
dash-dotted (-$.$) lines.
Statistical values: 
$\sigma^{2}_{x}=\sigma^{2}_{y}=2.0,\,  \tilde{\mu}_{x}=5.8\times10^{-2}, 
\,\tilde{\mu}_{y}=1.3\times10^{-2}$.
}
\label{fig3}
\end{figure}

In conclusion, we have shown (under realistic experimental conditions)
how phase imprinting of a staggered array of 
straight non-intersecting vortices in a harmonically trapped 
ultra--cold atomic gas can be used to generate and study quantum
turbulence in atomic BEC's.  
Finding velocity statistics similar to $^4$He 
in these relatively small systems means that atomic BEC's can be
used in the study of turbulence.
We have also found that the decay of turbulence is faster in the
presence of a thermal cloud, which we have modelled
in a simple way.

Our main result is that the statistics of the turbulent 
superfluid velocity
components are non--Gaussian, and exhibit power law tails similar
to what has been recently found in an experiment with
turbulent superfluid $^4$He \cite{Paoletti}.
We confirmed our 3D result by repeating the calculation in a 2D
condensate with a greater number of vortices, but lower 
dimensionless vortex line density,  in
a system in which vortex reconnections played virtually no role. 
Despite the huge difference in the ratio of
intervortex spacing to vortex core radius in atomic BEC's ($\approx 3$ here)
and in $^4$He ($10^5$ to $10^6$), 
we observed the same non--Gaussianity of the velocity, in
contrasts with the known Gaussianity of the velocity
in classical turbulence. Our result thus suggest that power-law
tails are a general feature of the constrained $1/r$ velocity fields which 
characterizes a quantum fluid.

The statistics of velocity
components are thus  macroscopic observables which
distinguish between classical and quantum turbulence. 
It is worth remarking \cite{pressure} that
another such observable is the pressure spectrum,
which, in 3D quantum turbulence, due to the $1/r$ velocity field 
and the singular nature of the vorticity, should obey a 
$k^{-2}$ law (where $k$ is the magnitude of the wavevector)
rather than the $k^{-7/3}$ scaling which corresponds to the classical
Kolmogorov $k^{-5/3}$ spectrum of the energy.

This research was supported by EPSRC grant EP/D040892/1 and by
the Merit Allocation Scheme of the National Facility of the 
Australian Partnership for Advanced Computing.


\begin{thebibliography}{99}
\bibitem{turbo-he4}
Walmsley, P. M. and Golov, A. I.,
Phys. Rev. Lett. {\bf 100}, 245301 (2008).

\bibitem{turbo-he3}
D. I. Bradley \etalcc, 
Phys. Rev. Lett. {\bf 101}, 065302 (2008);
V. B. Eltsov \etalcc, 
Phys. Rev. Lett. {\bf 99}, 265301 (2007).

\bibitem{Horng09}
T.L. Horng, C.H. Hsueh and S.C. Gou, 
Phys. Rev. A, {\bf 77}, 063625 (2008).

\bibitem{Tsubota-review}
K. Kasamatsu and M. Tsubota, in {\it Progress in Low Temperature
Physics} XVI, ed. by W.P. Halperin and M. Tsubota, pg. 351, Elsevier (2009).

\bibitem{Bagnato}  E. A. Henn {\it et al.},
Phys. Rev. Lett. {\bf 103}, 045301 (2009).

\bibitem{spectrum}
J.~Maurer and P.~Tabeling,
Europhys. Lett. {\bf 43}, 29 (1998);
C.~Nore, M.~Abid, and M.-E.~Brachet,
Phys. Rev. Lett. {\bf 78}, 3896  (1997);
T.~Araki, M.~Tsubota, and S.~K.~ Nemirovskii,
Phys. Rev. Lett. {\bf 89}, 145301 (2002);
M.~Kobayashi and M.~Tsubota,
Phys. Rev. Lett. {\bf 94}, 065302 (2005).

\bibitem{decay}
M.~R.~Smith \etalcc, 
Phys. Rev. Lett. {\bf 71}, 2583 (1993);
C.F. Barenghi and L. Skrbek,
J. Low Temp. Physics {\bf 146}, 5 (2007).

\bibitem{pipe}
P.~L.~Walstrom \etalcc, 
Cryogenics {\bf 28}, 101 (1998).

\bibitem{sphere}
M.~R.~Smith, D.~K.~Hilton, and S.~V.~Van~Sciver,
Phys. Fluids, {\bf 11}, 751 (1999).

\bibitem{classical-quantum}
W.~F.~ Vinen and J.~J.~ Niemela,
J. Low Temp. Phys. {\bf 128}, 167 (2002),
and Erratum, {\bf 129}, 213 (2002);
C.F. Barenghi, Physica D, {\bf 237}, 2195 (2008).

\bibitem{Paoletti}
M.S. Paoletti \etalcc, 
Phys. Rev. Lett.  {\bf 101}, 154501 (2008).

\bibitem{createturb}
N.G. Berloff and B. Svistunov,
Phys. Rev. A. {\bf 66}, 013603 (2002);
N.G. Parker and C.S. Adams,
Phys. Rev. Lett {\bf 95}, 145301 (2005);
M. Kobayashi and M. Tsubota,
Phys. Rev. A {\bf 76}, 045603 (2007);
M. Kobayashi and M. Tsubota,
J. Low. Temp. Phys. {\bf 150}, 587 (2008);
E. A. Henn \etalcc, 
Phys. Rev. Lett. {\bf 103}, 045301 (2009).

 
\bibitem{phase-imprinting}
S. Burger \etalcc, 
Phys. Rev. Lett., {\bf 83}, 5198, 1999;
A.E. Leanhardt \etalcc, 
Phys. Rev. Lett. {\bf 89}, 190403 (2002).

\bibitem{gamma_origin}
A.A. Penckwitt, R.J. Ballagh and C.W. Gardiner, Phys. Rev. Lett. {\bf 89}, 260402 (2002);
N.P. Proukakis and B. Jackson, J. Phys. B {\bf 41}, 203002 (2008).

\bibitem{gamma}
S. Choi, S.A. Morgan and K. Burnett, 
Phys. Rev. A. {\bf 57},  4057 (1998);
E.J.M. Madarassy and C.F. Barenghi,
J. Low Temp. Physics {\bf 152}, 122 (2008);
M. Tsubota \etalcc,
Phys. Rev. A {\bf 65}, 023603 (2002).
%

\bibitem{classical-turbo}
A. Vincent and M. Meneguzzi,
J. Fluids Mech. {\bf 225}, 1 (1991);
A. Noullez \etalcc, 
J. Fluid Mech. {\bf 339}, 287 (1997);
T. Gotoh \etalcc,
Phys. of Fluids {\bf 14}, 1065 (2002).

\bibitem{XMDSdoc}
Refer to http://www.xmds.org for a detailed description.

\bibitem{vortices-sound}
C.F. Barenghi \etalcc, 
J. Low Temp. Physics, {\bf 138}, 629 (2005);
M. Leadbeater \etalcc, 
Phys. Rev. A {\bf 67}, 015601 (2002).

\bibitem{Thesis}
Peter Mason, Ph.D. Thesis, University of Cambridge (2009).

\bibitem{reconnection-rate}
C.F. Barenghi and D.C. Samuels,
J. Low Temp. Physics {\bf 36}, 281 (2004).
M. Tsubota, T. Araki and S.K. Nemirowskii,
Phys. Rev. B {\bf 62}, 11751 (2000).

\bibitem{pressure}
 D. Kivotides \etalcc, 
Phys. Rev. Lett. {\bf 87}, 275302 (2001).





\end{thebibliography}
\end{document}